\documentclass[final,5p,times]{elsarticle}

\usepackage{amssymb}

\usepackage{array}
\usepackage{graphicx}
\usepackage{pstricks}
\usepackage{color}
\usepackage{xcolor}

\begin{document}

\begin{frontmatter}



\title{Neutrino oscillations in the presence of the crust
magnetization\tnoteref{t1}}
\tnotetext[t1]{Published in: Nuclear Instruments and Methods in Physics Research A 630 (2011) 242-245,
doi:10.1016/j.nima.2010.06.076.}



\author{J.~Syska\corref{cor1}
}
\ead{jacek.syska@us.edu.pl}


\address{Department of Field Theory and Particle Physics,
Institute of Physics, University of Silesia, Katowice, Poland}

\begin{abstract}
It is noted that the crustal magnetic spectrum exhibits the signal
from the partly correlated domain dipoles on the space-scale up to
approximately 500 km. This suggests the nonzero correlation among
the dynamical variables of the ferromagnetic magnetization
phenomenon on the small domain scale inside the earth's crust
also. Therefore the influence of the mean of the zero component of
the polarization on the CP matter-induced violation indexes is
discussed.

\end{abstract}


\end{frontmatter}



\section{Introduction: Magnetization of the lithosphere}
\label{magnetization in the lithosphere}

The nonzero magnetization of the geological structures of the
crust of the earth was reported some time ago in the geophysical
publications \cite{magnetization,Purucker-Mandea}. Hence the neutrino
oscillation phenomenon in the presence of the magnetization of the
earth's lithosphere is presented.  The geomagnetic analysis
\cite{Lowes-1966-1974-Langel-1987-Cain-1989,Cain-1990-Voorhies}
says that, due to the magnetized crust which possesses the induced
and remanent magnetization of the ferromagnetic origin, the
spacial magnetic field power spectrum differs from the core
exponential form. But the main inference from the statistical
analysis is that the dominant part of the crustal magnetic
spectrum has the intermediate form, which is
expected from the partly correlated domain dipoles on the
intermediate space-scale (up to approximately 500 km) of the
coherently magnetized geological structures
\cite{Cain-1990-Voorhies}.  It is the signal of the
(anti) parallel correlations of the spins of the ferromagnetic
(mainly iron's) domains on the far longer scale than the exchange
energy is able to explain, which in tern suggests the initial
condition for the analysis i.e. that they were formed as such.

\section{The effective $\nu$SM Hamiltonian and $\nu$SM transition rate}
\label{effective Hamiltonian}

The low-energy effective Standard Model ($\nu SM$) potential
Hamiltonian for the Dirac neutrino~($D$) charged current
interaction with the electrons of the medium has the following
form:
\begin{eqnarray} \label{H0D} \!\!\!\!\!\!\!\!\!\! ({\cal{H}}_{- -}^{0 \, D})_{ij} =
\sqrt{2} \ G_{F}  N_{e} \ (A_{LL}^{e})_{ij}  \left( \langle
\frac{\pi_{e}^{\mu}}{E_{e}} \rangle -  \langle \frac{m_{e} \,
s_{e}^{\mu}}{E_e} \rangle \right) \, n_{\mu}  \, ,
\end{eqnarray}
where the first and second term originate in the vector ($V$) and
axial-vector ($AV$) currents, respectively  \cite{AZS,ZZS}.
Indexes $i, j = 1,2,3$ are for three massive neutrinos $\nu_{i}$
and index $e$ stands for the background electrons. The quantities
$N_{e}$, $E_{e}$ and $m_{e}$ are the electron density, energy and
mass, respectively. The space component $\overrightarrow{n}$
of the four-vector $n^{\mu} = (1,\overrightarrow{n})$ points to
the direction of the neutrino momentum. $(A^{e}_{LL})_{ij} =
|\varepsilon_L^C|^2 \;U_{ei}^{L \ast}\; U_{ej}^{L}$ is the $\nu$SM
coupling \cite{Giunti Kim}, where $U^{L}$ is the unitary neutrino
mixing matrix in the charged (C) current interaction, and the
$V-AV$ factor $\varepsilon_{L}^{C}$ is the global $\nu SM$
coupling constant.

The mechanical four-momentum $\pi^{\mu}_{e}
\equiv (\pi_{e}^{0}, \vec{\pi}_{e})$ is defined as $\pi^{\mu}_{e}
= p^{\mu}_{e} - e A^{\mu}$, where $A^{\mu}$ is the electromagnetic
four-potential acting on the background electron. The electron
polarization four-vector $s_{e}^{\mu}$ is equal to:
\begin{eqnarray}
\label{S-more general} s_{e}^{\mu} = \left[ \frac{{\vec{\pi}}_{e}
\cdot {\vec{\lambda}_{e}}}{m_{e}}, \; {\vec{\lambda}_{e}} +
\frac{{\vec{\pi}}_{e} ( {\vec{\pi}}_{e} \cdot {\vec{\lambda}_{e}}
) }{m_{e} (m_{e} + E_{e})} \right] \equiv (s_{e}^{0}, \vec{s}_{e})
\; ,
\end{eqnarray}
where ${\vec \lambda}_{e} = \chi_{e}^{\dag} {\vec \sigma}
\chi_{e}$ ($\, \chi_{e}^{\dag} \chi_{e} = 1$) is the electron's
polarization and $\chi_{e}$ is its two component spinor. The
quantities $\langle \frac{\pi^{\mu}_{e}}{E_{e}} \rangle$ and $
\langle \frac{m_{e}\, s_{e}^{\mu}}{E_{e}} \rangle$ are the
thermodynamical means. Finally, the $\nu SM$ meets both the
relation $({\cal{H}}_{+ +}^{0 \, D})_{ij} = 0$ and $({\cal{H}}_{+
+}^{0 \, \overline{D}})_{ik} =
 - ({\cal{H}}_{ -  \ - }^{0 \, D})^{*}_{ik}$ in the case of the  Dirac
antineutrino ($\overline{D}$).

\subsection{The thermodynamical means}
\label{The thermodynamical means}

With the isotropy assumption  of the electron momentum
\cite{Giunti Kim}, $\langle \vec{\pi}_{e} \rangle \approx 0$, we
obtain $\langle \frac{\pi^{\mu}_{e}}{E_{e}} \rangle \equiv (
\langle \frac{\pi^{0}_{e}}{E_{e}} \rangle \ , \langle
\frac{\vec{\pi}_{e}}{E_{e}} \rangle ) \approx ( 1 \ , \vec{0} \ )$
for the $V$ term in Eq.(\ref{H0D}). From Eq.(\ref{S-more general})
wee see that for the $AV$ term in Eq.(\ref{H0D}) the crust
unpolarized matter condition $ \langle \vec{s}_{e} \rangle \approx
0$ leads to $\langle \frac{m_{e} \, s_{e}^{\mu}}{E_{e}} \rangle
\approx \left( \langle \frac{{\vec{\pi}}_{e} \cdot
{\vec{\lambda}_{e}}}{E_{e}} \rangle \ , \vec{0} \ \right)$.

As besides the initial condition mentioned in Section~1, the crust
impact on the correlations seen in the magnetic power spectrum
around the globe has mainly the ferromagnetic origin
\cite{magnetization} hence for the temperatures in the crust
\cite{book_geomagnetism} the magnetic moments inside each of the
ferromagnetic domains are overwhelmingly parallel arranged
\cite{Patterson_Bailey}. Due to the potential $A^{\mu}$ inside one
domain there appears the nonzero mean correlation between
polarization ${\vec \lambda}_{e}$ and mechanical momentum
${\vec{\pi}}_{e}$ inside each of the ferromagnetic domains, having
the same sign for every one of them. It results in the nonzero
mean value of the zero polarization component $\langle s_{e}^{0}
\rangle \neq 0$ along the whole experimental baseline $L$ (which
stands in the fundamental opposition to the earth's crust mean
space component behavior $\langle \vec{s}_{e} \rangle \approx
\vec{0} \, $). The full solid-state analysis should follow. To
test the impact of the described phenomenon on the neutrino
oscillation in the crust the natural baselines of $L \leq 874$ km
for the current experiments could be used. But the specially
builded plants with the short but totaly ferromagnetic baselines
are thinkable also.

\subsection{The $\aleph_{e}$ magnetization form of the potential
Hamiltonians and transition rate formula}

The above considerations lead to the Dirac neutrino
and antineutrino $\nu$SM Hamiltonians:
\begin{eqnarray}
\label{H0D_wycechowane}  & & ({\cal{H}}_{- -}^{0 \, D})_{ij} =
\sqrt{2} \ G_F   N_{e} (A_{LL}^{e} )_{ij} ( 1 - \aleph_{e} ) \; ,
\nonumber
\\ & & ({\cal{H}}_{+ +}^{0 \, \overline{D}})_{ij} = -
({\cal{H}}_{- -}^{0 \, D})^{*}_{ij} \; ,
\end{eqnarray}
where
\begin{eqnarray}
\label{alef_e} \aleph_{e} \equiv \langle \frac{{\vec{\pi}}_{e}
\cdot {\vec{\lambda}_{e}}}{E_{e}} \rangle \, .
\end{eqnarray}
Here $\aleph_{e}$ is the only nonzero term connected with the
magnetization of the background electrons. The $\nu$SM effective
Hamiltonians ${\cal{H}}$ for the neutrino or antineutrino are, in
the helicity - mass base $|\lambda, i \rangle$, $\lambda = \pm
\frac{1}{2}, \; i = 1,2,3$, the $3\times3$ dimensional matrices:
\begin{eqnarray}
\label{decomp H} & & {\cal{H}}^{D} = {\cal M} + {\cal{H}}_{--}^{0
\, D} \; , \;\;\; {\cal{H}}^{\overline{D}} =
{\cal M} + {\cal{H}}_{++}^{0 \, \overline{D}} \; , \nonumber \\
& & {\cal M} = diag(E_{1}^{0}, E_{2}^{0}, E_{3}^{0}) \; .
\end{eqnarray}
${\cal M}$ is the vacuum mass term, $ E_{i}^{0} = E_{\nu} +
\frac{m_{i}^{2}}{2 E_{\nu}}$,  and $E_{\nu}$ is the energy for the
massless neutrino \cite{Giunti Kim}. In the $\nu$SM and for the
homogenous medium and (in practise) relativistic neutrinos, the
$\alpha$ to $\beta$ flavor oscillation probability $P_{\alpha
\rightarrow \beta}(L) $ factors out in the differential transition
rate formula \cite{ZZS}:
\begin{eqnarray} \label{transition cross sec nuSM relat}
\frac{d \sigma_{\beta \; \alpha}}{d \Omega_{\beta}} &=& f_{D} \,
A^{L}_{\varepsilon \varepsilon} \, \sum_{i; \, i'} \, U^{L}_{\beta
i} U^{L \ast}_{\beta i'} U^{L \ast}_{\alpha i} U^{L}_{\alpha i'}
\, e^{\; i \, \Delta m^{eff \, 2}_{\ i \, i'}  \, L/(2 \,
E_{\nu})} \nonumber \\ &\equiv& \frac{d \sigma_{\beta}}{d
\Omega_{\beta}}(m_{i}=0) \times P_{\alpha \rightarrow \beta} (L)
\, ,
\end{eqnarray}
where $f_{D}$ is the kinematical factor, $A^{L}_{\varepsilon
\varepsilon}$ is the function of the energies and momenta of the
particles in the detection process, and $\Delta m^{eff \, 2}_{\ i
i'}$ is the neutrinos effective square mass difference in the
medium calculated with Eqs.(\ref{H0D_wycechowane}) - (\ref{decomp
H}).


\section{The numerical results. Advancing steps in the analysis}
\label{numerical results}

The variety of neutrino oscillation observables could be used for
the purpose of the vacuum oscillation parameters estimation. As
the experiments are performed on the earth hence the dependance
of these observables on the crust magnetization has to be well
understood. The simplest one is the oscillation probability
$P_{\alpha \rightarrow \beta} (L)$ plotted on Figure~1 up to the
limit baseline $L=874$ km (taken as the approximate maximum value
of the neutrino path in the earth's crust).
\begin{figure}[here]
\includegraphics[angle=0,width=70mm]{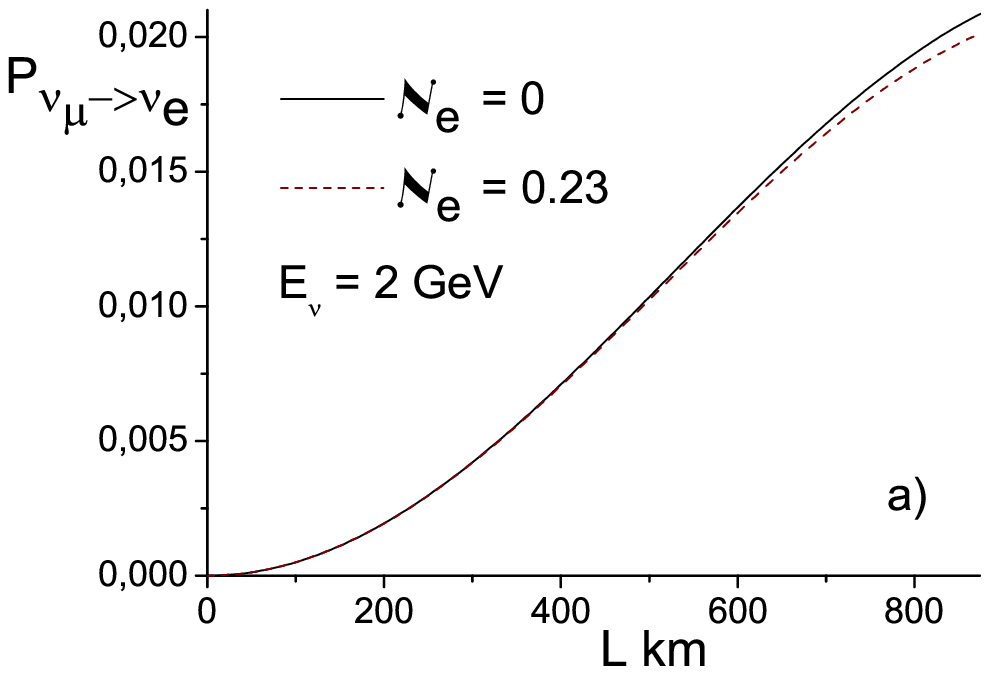}
\includegraphics[angle=0,width=70mm]{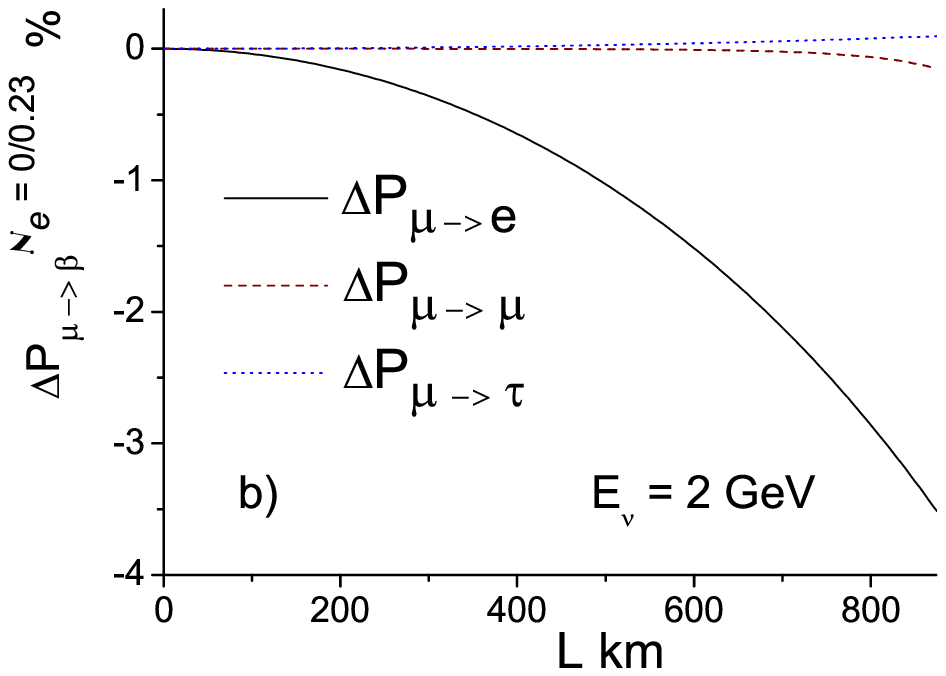}
\caption{Panel: (a)~The (noticeable) change of $P_{\nu_\mu
\rightarrow \nu_{e}}$ with $\aleph_{e}$. (b)~The relative change
of $\Delta P_{\mu \rightarrow \beta}^{0/\aleph_{e}} \equiv 100 \%
\,(P_{\mu \rightarrow \beta}^{\aleph_{e}} - P_{\mu \rightarrow
\beta}^{0})/P_{\mu \rightarrow \beta}^{0}$, ($\beta = e, \mu,
\tau$), with $\aleph_{e}$ is even more visible. The curves for
$\aleph_{e} = 0$ vs. $\aleph_{e} = 0.23$ are plotted. The neutrino
energy is taken to be $E_{\nu} = 2$ GeV.} \label{twosubfigures}
\end{figure}
Then the transition rate given by Eq.(\ref{transition cross sec
nuSM relat}) for the number of events in the detector follows.
Yet, in the full oscillation data analyzes the number of the
events is useful for the preliminary analyzes only. What matters
is the functional dependance of the observable on the probability
oscillation in the particular phenomena also. The steps from the
observable unsensitive to the $\aleph_{e}$ magnetization to the
sensitive ones are as follows:
\begin{enumerate}[1.]
\item The observable unsensitive to the $\aleph_{e}$ magnetization
of the earth's crust is $R_{\mu/e}$, the ratio-of-ratios for the
muon vs. electron atmospheric neutrinos \cite{Giunti Kim}. Its
weak dependance on $\aleph_{e}$ is connected with the general fact
that the linear dependance of the transition rates on ${\cal{H}}_{
-  -}^{0 \, D}$ and ${\cal{H}}_{ + +}^{0 \, \overline{D}}$ cancels
out (due to opposite signs for $D$ and $\overline{D}$ in
Eq.(\ref{H0D_wycechowane})). Hence $R_{\mu/e}$ would be perfect
for the vacuum parameters estimation. Unfortunately the
experimental errors for $R_{\mu/e}$ are bigger than 3\%. \item The
observables dependent on the $\aleph_{e}$ magnetization:
\begin{enumerate}
\item The up-down asymmetry $A^{up-down}_{\alpha}$ in the
atmospheric neutrinos experiments \cite{Giunti Kim} seems to be
unprofitable for the decision about the solitary earth's crust
$\aleph_{e}$ dependance.
Therefore the following paper is going to be devoted to the
all-directions analyzes on $\aleph_{e}$ dependance in the
'through-earth' up-down asymmetry. \item The CP matter-induced
violation $A^{CP}_{\alpha \rightarrow \beta}$ is sensitive to
$\aleph_{e}$ (see Section~\ref{sensitive to alef}). \item The
$\aleph_{e}$-CP matter-induced asymmetry defined below is the
observable (very) sensitive to the $\aleph_{e}$ magnetization. The
accelerator neutrino could be taken into account also.
\end{enumerate}
\end{enumerate}

\subsection{The observable sensitive to magnetization: \\
The CP matter-induced violation} \label{sensitive to alef}

Even when the fundamental Lagrangian is CP symmetric (the mixing
matrix $U^{L}$ phase $\delta=0$), we could observe the
matter-induced violation of the CP symmetry expressed in the
(non-vanishing) difference \cite{Giunti Kim}:
\begin{eqnarray}
\label{CP asymmetry} A^{CP}_{\alpha \rightarrow \beta} =
P_{\nu_{\alpha} \rightarrow \nu_{\beta}} -
P_{\overline{\nu}_{\alpha} \rightarrow \overline{\nu}_{\beta}} \;
, \;\;\; \alpha, \beta = e, \, \mu, \, \tau \; .
\end{eqnarray}
The plots on Figure~2 show, for relatively low neutrino energy
($E_{\nu} = 2$ GeV) and $L$ = 500 or 874 km, the noticeable change
of $A^{CP}_{\mu \rightarrow \beta}$ with $\aleph_{e}$.
\begin{figure}[here]
\vspace{-2mm}
\includegraphics[angle=0,width=72mm]{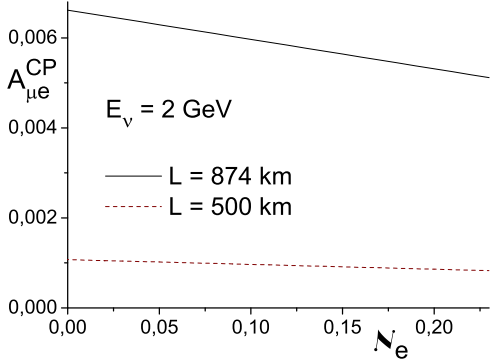}
\vspace{-2mm}
\includegraphics[angle=0,width=72mm]{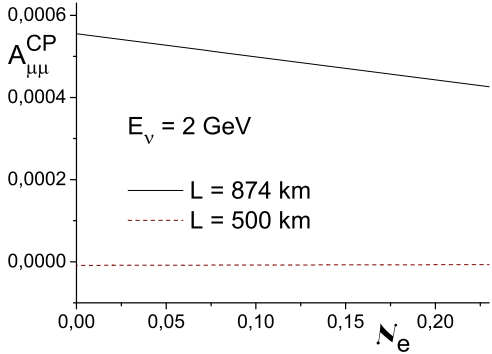}
\vspace{-2mm}
\includegraphics[angle=0,width=72mm]{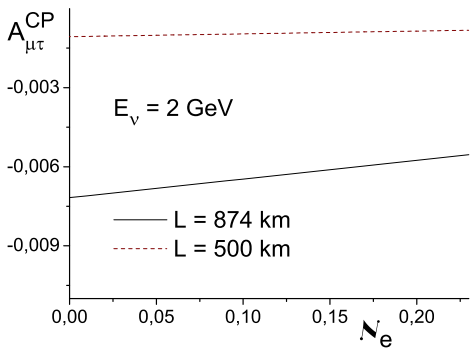}
\vspace{-2mm}
\caption{The change of $A^{CP}_{\mu \rightarrow \beta}$ with
$\aleph_{e}$ for $\beta = e, \mu, \tau$ (from top to bottom) and
for the neutrino energy $E_{\nu} = 2$ GeV and $L$ = 500 or 874
km.
\vspace{4mm}}
\end{figure}
Hence with $A^{CP}_{\alpha \rightarrow \beta}$ we have finally
obtained the observable sensitive to the earth's crust
$\aleph_{e}$ magnetization, although, because of the experiment
sensitivity to $\Delta m^2$, with the increase of $E_{\nu}$ the
$\aleph_{e}$ dependence is decreasing.

\subsection{The $\aleph_{e}$-CP matter-induced asymmetry. Even the
accelerator neutrinos matter} \label{aleph-CP matter-induced
asymmetry}

The $\aleph_{e}$-CP mater-induced asymmetry could be defined as
follows:
\begin{eqnarray}
\label{aleph CP asymmetry} \!\!\!\!\!\!\!\!\!\!\!\!\!
A^{\aleph_{e}-CP}_{\alpha \rightarrow \beta} =
\sigma_{\nu_{\alpha} \rightarrow \nu_{\beta}} -
\sigma_{\overline{\nu}_{\alpha} \rightarrow
\overline{\nu}_{\beta}} = \sigma_{N \, \nu_{\beta}}
P_{\nu_{\alpha} \rightarrow \nu_{\beta}} - \sigma_{N \,
\overline{\nu}_{\beta}}  P_{\, \overline{\nu}_{\alpha} \rightarrow
\overline{\nu}_{\beta}} \; ,
\end{eqnarray}
where the initial flavors are $\alpha = e, \mu$ or $\tau$ and
index $\beta = e, \mu$ or $\tau$ is for the events in the
detector. The second equality in Eq.(\ref{aleph CP asymmetry}) is
valid for the $\nu$SM relativistic case (see Eq.(\ref{transition
cross sec nuSM relat})), where the index $N$ in the total cross
sections\footnote{For information
on
$\sigma_{N \, \nu_{\beta}}$ and $\sigma_{N \,
\overline{\nu}_{\beta}}$ see:
J.A.~Formaggio, G.P.~Zeller,
From eV to EeV: Neutrino cross sections across energy scales,
Rev.Mod.Phys.~84 (2012) 1307-1341,  arXiv:1305.7513 [hep-ex].
%
}
$\sigma_{N \, \nu_{\beta}}$ and $\sigma_{N \,
\overline{\nu}_{\beta}}$ signifies the
nucleon.
Let us notice that
even if $A^{CP}_{\alpha \rightarrow \beta} = 0$ we have
$A^{\aleph_{e}-CP}_{\alpha \rightarrow \beta} \neq 0$.  We could
define also the ratio:
\begin{eqnarray}
\label{aleph R CP asymmetry} RA^{\aleph_{e}-CP}_{\alpha
\rightarrow \beta} = 100 \% \, \frac{A^{\aleph_{e}-CP}_{\alpha
\rightarrow \beta}(\aleph_{e}) - A^{\aleph_{e}-CP}_{\alpha
\rightarrow \beta}(\aleph_{e}=0) }{ A^{\aleph_{e}-CP}_{\alpha
\rightarrow \beta}(\aleph_{e}=0)} \, .
\end{eqnarray}
Let us make some predictions e.g. for the K2K experiment.  Here
for $L = 250$ km and $E_{\nu} = 1.3$ GeV the neutrino beam was the
almost pure muon neutrino one. Hence to calculate
$A^{\aleph_{e}-CP}_{\mu \rightarrow \beta}$ the experiment should
be rebuild for the possibility of the production of the
${\overline \nu}_{\mu}$ beam also. Then with $\aleph_{e}$ = 0.23
we obtain $RA^{\aleph_{e}-CP}_{\mu \rightarrow e} \approx $ $-20$\%
which validates the conclusion that with $A^{\aleph_{e}-CP}_{\mu
\rightarrow \beta}$ the correction of the experimental values of
$\sigma_{N \, \nu_{\beta}}$ for the neutrino-nucleon inelastic
scattering would be possible also.

\vspace{-2mm}

\section{The conclusions}
\label{conclusions}

\vspace{-2mm}

The impact of the zero component of the mean
polarization $\langle s_{e}^{0} \rangle$ from ferromagnetic
domains of the earth's crust on the basic neutrino oscillation
observables has been discussed. Two of them, i.e. the CP
matter-induced violations, $A^{CP}_{\alpha \rightarrow \beta}$ and
$A^{\aleph_{e}-CP}_{\alpha \rightarrow \beta}$, exhibit the
greatest influence of the $\aleph_{e} \equiv \langle
{\vec{\pi}}_{e} \cdot {\vec{\lambda}_{e}}/E_{e} \rangle$
background media magnetization. It follows that under the precise
knowledge of the $\aleph_{e}$ magnetization along the baseline $L$
(natural or artificially designed) and for relatively low neutrino
energies (2 - 10 GeV or even for the reactor's ones) we could
improve both the estimation of the vacuum neutrino oscillation
parameters\footnote{
Or that
to improve the estimation of the vacuum neutrino oscillation parameters, the precise knowledge of $\aleph_{e}$ along the baseline $L$ is necessary.
}
and the neutrino cross sections also. Finally, as for
the purpose of the geomagnetic analysis of the huge Minto
block\footnote{Its northern part is located approximately 2400 km from Fermilab.}
in Canada \cite{Pilkington_Percival} the samples from the rocks were
taken every 10~km, hence the direct measurements of the crust
magnetization for one e.g. 295 km long neutrino T2K
baseline should not be the too difficult task as well.
\\

\vspace{-2mm}

\hspace{-4mm}{\bf Acknowledgments}: This work has been supported by L.J.Ch..\\
This work has been also supported by the Junta de Andaluc{\' i}a
project FQM project FQM 437 and by the Institute of Physics,
University of Silesia, Poland.
Special thanks to F.~del~Aguila, J.~Aksamit,
J.~Holeczek, J.~Kisiel, M.~Mierzejewski, R.~Szafron, S.~Zaj{\c a}c
and M.~Zra{\l}ek for the discussions. I would also like to extend my thanks to
Heptools for its support.



\bibliographystyle{elsarticle-num}
\bibliography{<your-bib-database>}

\end{document}